*Measurement of trace elements in some popular medicinal plants using Instrumental Neutron Activation Analysis method (INAA) in Arak, Iran*

*R. Pourimani*[*1]; S. Kashian[2]; A.A. Fathivand[3]*

[1] *Department of Physics, Faculty of Science, Arak University, Arak, Iran*
[2] *Radiation Applications Research School, Nuclear Science and Technology Research Institute, Tehran, Iran*
*\* Corresponding author email address: r-pourimani@araku.ac.ir   ORCID: 0000-0002-0102-0578*

**Abstract-** *In this work, the specific mass of twelve elements were determined in five of the most commonly used medicinal plants as Caraway (Carum carvi), Savory (Satureia hortensis), Purslane (Portulaca oleracea), Fenugreek (Trigonella foenum-graecum) and Milk thistle (Silibum marianum) prepared from herbal pharmacies. Multi elemental Instrumental Neutron Activation Analysis (INAA) method was applied to analyze the samples. Tehran research reactor was used as a neutron source and gamma ray spectra registered using high purity germanium (HPGe) detector. Among analyzed samples, highest concentrations of Fe (8789 ppm), Cr (8 ppm) and Na (517 ppm) were found in Caraway. Maximum levels of Mn (95 ppm), Cl (3702 ppm), Ca (18328 ppm) , K (21562 ppm) and V (2.7 ppm) were detected in Savory and Fenugreek contains the lowest concentrations of Fe (195 ppm), Zn (13 ppm), Ca (2243 ppm), Al (99ppm), Mn (26 pm) and Mg (177ppm).*

*INTRODUCTION*

*Currently, many people are interested in using medicinal herbs to improve the disease, and their interest is growing every day. That is why it is important to determine the content of useful and unprofitable elements in them. Many research have been carried out on the benefits of herbal medicine and many articles have been published, such as medicinal plants used to treat neuritis colitis [1], some herbs used to treat skin diseases [2] herbs bring down blood cholesterol [3], plants that bring down blood glucose [4] and Indian herbs are used to control diabetes [5]. The most important research and articles relate to the way the herbs work on the body to improve the disease. In addition, researchers have been done to determine the concentrations of the elements in the medicinal plants, which can be noted: Application of the neutron activation analysis (NAA) method for the determination of trace elements in six medicinal plants in Ethiopia [6] determination of the essential elements in Indian medicinal plant [7], assessment of low levels of elements in Chinese herbs [8], metal determination in Brazilian medicinal plants [9]. The main purpose of this work was to determine the specific mass of Ca, Cr and Mg, which can be very effective in the prevention and treatment of lipid disorders. [10]. In this study five kinds of popular medicinal herbs that are widely used in Arak in the treatment of blood lipid disorders such as succory, anet dill seed, thyme, fumitory and sorrel karkade were analyzed using Instrumental Neutron Activation Analysis (INAA) technique. Also has been determined the specific mass of Fe, Zn, Br, V, K, Na, Cl, Al, and Mn, which is an important source of metabolic functions. This method is a suitable technique, because samples can be analyzed simultaneously for a large number of elements with more precise measurement results. In addition, it is a non-destructive measurement method and there is no need to digest the sample, and only a little quantity are needed. In contrast to chemical drugs, medicinal plants contain a large number of active substances, all of which are consistent, with the nature of the human body. Among the herbs used to lower blood fat are caraway, Savory, Purslane, Fenugreek seed and Milk thistle. Knowing the ingredients that make up these herbs and the drug how can affect the treatment.*

*Experimental method*
*Sampling and sample preparation*

*In this study, five popular herbal plants samples as Caraway (Carum carvi), Savory (Satureia hortensis), Purslane (Portulaca oleracea), Fenugreek seed (Trigonella foenum-graecum seed) and Milk thistle (Silibum marianum) were collected from Arak herbal pharmacies in Iran. All samples were washed using by twice distilled water. They were kept at room temperature for two days. The samples were then kept at an oven at 80 °C temperature until reached to constant weight. After drying, the samples were chopped, milled and weighted. Then, samples were changed into powders using an agate mortar. In order to make them homogenous, sieved to pass through mesh with 0.149 mm diameter. Samples ( about 300-350 mg) were sealed in polyethylene vial prior to neutron irradiation. Four samples of each type of herbal plant were prepared and analyzed and were determined average and standard deviation. Similarly, for quality control and standard samples used reference*

materials IAEA-V-10 and IAEA-336 respectively, which are similar to samples were irradiated and were analyzed together with samples. Using the pneumatic sample transfer system, the sample and standard reference material (IAEA –336) were irradiated for 2 min at Tehran Research Reactor. The thermal neutron flux was about $3\times10^{13}$ n.cm$^{-2}$.s$^{-1}$ [11]. The sample and standard reference material were counted immediately after irradiation to determine Al and were counted 15 min after irradiation to determine the amount of elements with short half-life of radionuclides such as Mn, Na, Mg, Ca, Cl and K. For measurement amount of Fe, Zn, Cr, and Br, the samples were irradiated for 2 h and counted after one week. Half life time of isotopes was taken from [12]. The radionuclides used in the analysis with their gamma energies, irradiation time, and counting time and conditions are listed in Table 1.

*Table 1. Radionuclides, gamma ray energies, irradiation and counting time and conditions*

| Element | Radioisotpe | Half life | Energy (keV) | Irradiation and counting conditions |
|---------|-------------|-----------|--------------|-------------------------------------|
| Al      | $^{28}$Al   | 2.24 m    | 1779         | 2 min irradiation, counted immediately after irradiation for 300 s |
| Mn      | $^{56}$Mn   | 2.58 h    | 1810         | 2 min irradiation, counted 15 min after irradiation for 900 s |
| Na      | $^{24}$Na   | 15.00 h   | 1368         | |
| Mg      | $^{27}$Mg   | 9.46 m    | 1014         | |
| K       | $^{42}$K    | 12.40 h   | 1525         | |
| Cl      | $^{38}$Cl   | 37.24 m   | 1642         | |
| Ca      | $^{49}$Ca   | 8.72 m    | 3082         | |
| Fe      | $^{59}$Fe   | 44.63 d   | 1099         | 2 h irradiation, counted 1 week after irradiation for 60000 s |
| Zn      | $^{65}$Zn   | 244 d     | 1115         | |
| Cr      | $^{51}$Cr   | 27.69 d   | 320          | |
| Br      | $^{82}$Br   | 35.4 h    | 554          | |

*Spectrometry and spectrum analysis*

After irradiation, samples were placed for gamma ray spectrometry in face to face of high purity germanium (HPGe) detector. Detection was performed using (EGPC 5574 model, manufactured by Intertechnique, France) with 10% relative efficiency coupled with multi-channel analyzer. The energy resolution (FWHM) for gamma-ray ($^{60}$Co) with 1332.52 keV energy was 1.95 keV. The detector and preamplifier are shielded in a chamber of three layers included of 10 cm thick lead, 1 mm thick cadmium, and 2 mm thick by copper. This shield serves to reduce background radiation. The soft components of cosmic ray, consisting of photons and electrons, are reduced to a very low level by 100 mm of lead shielding. The X-ray (73.9 keV) emitted from lead by its interaction with external radiation is suppressed by copper layer and cadmium layer successively, absorbing thermal neutrons produced by cosmic ray [13].

The concentration of the elements in the activated samples were calculated quantitatively using the gamma-ray spectrum analysis software packages known as Winspan 2004 that uses equation (1) to determine the concentrations of the elements in a sample in comparison with the known masses of the elements in reference material.

$$C_s = C_{st} \frac{A_s \left(e^{-\lambda t_d}\right)_{st}}{A_{st} \left(e^{-\lambda t_d}\right)_s} \qquad (1)$$

Where, $C_s$ and $C_{st}$ are concentrations of unknown analyte (interested element) in the sample and standard; $A_s$ and $A_{st}$ are the activities of unknown analyte in the sample and standard, $t_d$ is the decay time of the unknown element in the sample and in the standard, $\left(e^{-\lambda t_d}\right)_s$ and $\left(e^{-\lambda t_d}\right)_{st}$ are the decaying factors for sample and standard sample respectively.

*The accuracy and precision of the measurement method was evaluated by analyzing the reference material IAEA-V-10 known as Hey powder. The qualities of obtained results for this analysis were evaluated using z-score method (equation 2) [14].*

$$z = \frac{x-c}{\sqrt{u_x^2 + u_c^2}} \qquad (2)$$

*Where x is the analytical result, c is the certified value, $u_x$, $u_c$ are uncertainties of measured and certified values. The uncertainty of the analytical results is based on counting statistics whilst the uncertainty of certified value is based on the certificate. However, the Horwitz function is used to estimate the uncertainties of information values in the certificates [15]. For acceptance of results: -2<z<2 is anticipated. However, if z<-3 or z>3, it is consider that the result is "out-of-control" and corrective action must be taken. As seen from table 2, for all interested elements the z-score value ranges within ±2 which indicates that measured concentrations of elements in the activated reference material were in good agreement with certified values.*

*Four samples of each kinds of medicinal herbs were irradiated and measured and for all samples were calculated standard deviation. The results has been shown in figure 1. From this figure it is observed that, all of the measurements are within the acceptance criteria.*

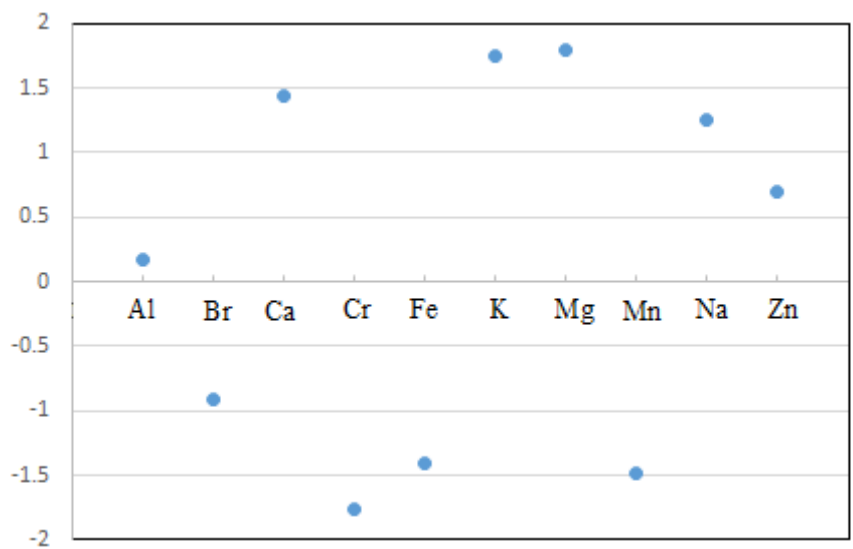

*Figure1- Control charts showing the z-scores of the analyzed reference materials IAEA-V-10.*

**Result and discussion**

*For Quality control purpose, the reference material IAEA-V-10 was prepared irradiated in same condition and gamma ray spectrum registered and was analyzed. Results of this analysis are listed in Table 2. Standard deviation and Z-score calculation shown in figure 1 which for all samples the Z-Score ranges within ±2. The specific mass of trace elements were calculated for all herbal plants samples. Result for average trace elements concentrations included common and scientific name are given in Table 3.*

***Table 3*** *Mean elemental concentrations and standard deviations medicinal plants (ppm)*

| Common | Caraway | Savory | Purslane | Fenugreek | Milk |
|---|---|---|---|---|---|

| Scientific name | Carum carvi | (Satureia hortensis), | (Portulaca oleracea), | seed (Trigonella foenum-graecum seed) | thistle (Silibum marianum) |
|---|---|---|---|---|---|
| Fe | 8789±55 | 1716±36 | 269±14 | 195±8 | 804±43 |
| Zn | 35±2 | 19±1 | 41±2 | 13±1 | 154±7 |
| Cr | 8.0±0.1 | ND | 1.5±0.1 | 1.3±0.1 | ND |
| Br | ND | ND | 3.0±0.1 | 5.0±0.1 | 2.0±0.1 |
| V | ND | 2.7±0.3 | 0.6±0.1 | ND | 1.9±0.1 |
| K | 12060±815 | 21562±862 | 9286±527 | 14533±581 | 8142±391 |
| Na | 517±9 | 439±5 | 127±3 | 535±13 | 497±7 |
| Ca | 11400±1036 | 18378±1318 | 2729±447 | 2243±224 | 12342±871 |
| Cl | 1518±29 | 3702±43 | 449±12 | 1990±29 | 3062±83 |
| Al | 559±27 | 753±23 | 131±10 | 99±14 | 1197±20 |
| Mn | 35±1 | 95±2 | 77±2 | 26.0±0.1 | 64±1 |
| Mg | 3409±311 | 2147±121 | 3915±826 | 177±12 | 2992±190 |

*Our result shows that the Caraway contain maximum concentrations of Cr (8 ppm), Fe (8789 ppm) and Na (517 ppm). Maximum concentration of Zn (154 ppm) obtained for Milk thistle and highest amount of Mg (3915 ppm) observed in Purslane sample. The Savory contains maximum concentration of Ca (18378ppm) and Mn (95 ppm). These herbal plants may be using an ample source of Cr, Zn, Mg and Ca for human. Our results confirmed that these herbal plants can helps people to reduce the level of lipids and cholesterols because contained elements such Ca, Cr and Mg [16]. Also most of these medicinal plants contain useful elements such Fe, K. Maximum amount of Br (5 ppm) observed in Fenugreek seed sample. Amount of Br advised by the International agriculture organization and international health organization is (0–20 mg/kg) for product which are not disinfection [17]. The Br tolerable daily intake for absorbed daily in body by FAO/WHO range of 0–1 mg/kg of body weight [17]. This value was cover this range and is the same range for wheat crops of Iran [18]. For the purpose of method validation and quality assurance, quality control sample IAEA-V-10 known as Hey powder was analyzed together with the samples. Results of the measurements show that measured concentrations of elements in the activated reference material are in good agreement with the certified values.*

*Conclusion*

*In this research, concentrations of twelve elements were determined in five medicinal plants employing instrumental neutron activation analysis method. According to this method were calculated amount of Fe, Zn, Br Cr, Br, V, K, Na, Al, Mn and in Caraway, Savory, Purslane, Fenugreek seed and Milk thistle. These medicinal plants are widely used by the population of Iran to prevent the various diseases. The results of this research indicated that Savory and Milk thistle contains highest levels of Ca as 18378 and 12342 ppm respectively. The highest amount of Cr and Mg obtained for Caraway (8 ppm) and Pursalne (3915 ppm) respectively. These elements have major role in reducing Cholesterol and Triglyceride levels in the blood. Results also show that studied medicinal plants are enriched in essential elements for the metabolic function.*

*Acknowledgements*

*The authors would like to thank Arak University Research Council and Nuclear Science and Technology Research Institute of Tehran for their financial support.*